\documentstyle[11pt]{article}

\begin{document}
\title{\bf A note on ``Weighted Evolving Networks: Coupling Topology and Weight Dynamics"}
\author{R.~V.~R.~Pandya \thanks{Email: {\sl rvrpturb@uprm.edu}} \\
Department of Mechanical Engineering, University of Puerto Rico at
Mayaguez,\\ Mayaguez, Puerto Rico, PR 00680, USA}

\date{\today}
\maketitle

{PACS number(s): 89.75.Hc, 05.40.-a, 87.23.Kg}

In a recent Letter \cite{BBV04}, Barrat,
Barth\'{e}lemy and Vespignani (BBV) have proposed a model for the
evolution of weighted network when new edges and vertices are
continuously established into the network while causing dynamic
behavior of the weights. Their model dynamics starts from some
initial number of vertices connected by links or edges with
assigned weights and at each time step, addition of a new vertex
$n$ with $m$ edges and subsequent modification in weights are
governed by the following two rules:
\begin{enumerate}
\item The vertex $n$ is attached at random to a previously
existing vertex $i$ according to the probability distribution
\begin{equation}
\Pi_{n\rightarrow i}=\frac{s_i}{\sum_{j}\,s_j}. \label{eq1}
\end{equation}
\item The induced total increase $\delta$ in strength $s_i$ of the
$i$th vertex is distributed among the weights $w_{ij}$ of its
neighbors $j$ according to
\begin{equation}
w_{ij}\rightarrow w_{ij}+ \delta \frac{w_{ij}}{s_i}. \label{eq2}
\end{equation}
\end{enumerate}
This second rule, though could be one possibility, does not follow
the same mechanism of the first rule.
Here we discuss these rules in the context of worldwide airport
network and suggest an alternative to the second rule which is
consistent with the mechanism of the first rule.

In BBV's own words, the first rule can be described as ``busy get
busier" \cite{BBV04b}. It can be written more explicitly as ``busy
airports get busier". The Eq. (\ref{eq1}) suggests that it is more
probable that a new airport (vertex) $n$ will be attached to the
airport $i$ which handles more traffic represented by strength
$s_i$. The second rule (Eq. \ref{eq2}) does not follow the same
mechanism, instead it can be described by ``busy routes get
busier". According to the second rule, the route $i$ to $j$ having
more traffic as indicated by $w_{ij}$ would handle larger portion
of the induced traffic $\delta$ given by
$\delta\frac{w_{ij}}{s_i}$. That does not necessarily mean that
the airport $j$, in the neighbor of $i$, with largest value for
$w_{ij}$ is also the airport with maximum strength or traffic in
comparison with other neighboring airports of $i$. Now, as an
alternative to Eq. (2), consider
\begin{equation}
w_{ij}\rightarrow w_{ij}+\delta\frac{s_j}{\sum_{k \in
{\cal{V}}(i)}\,s_k} \label{eq3}
\end{equation}
where ${\cal{V}}(i)$ indicates set of all neighboring airports
(vertices) of $i$ and $k\neq n$. The last term of Eq. (\ref{eq3})
indicates that it is more probable that the induced traffic would
go towards the airport $j$ which handles maximum traffic $s_j$
among the neighboring airports ${\cal{V}}(i)$ of $i$. Thus, this
mechanism is in consistency with the mechanism of the first rule,
i.e. busy airports get busier.

Also, it should be noted that the second rule of BBV does not
consider further redistribution of $\delta\frac{w_{ij}}{s_i}$
among the weights of the neighbors of neighbors of airport $i$.
And BBV's weighted model is limited to the case where passengers
prefer direct flights or/and flights with one connection. In order
to include the flights with two intermediate connections and in
accordance with the first rule, $\delta'\equiv
\delta{s_j}/[{\sum_{k \in {\cal{V}}(i)}\,s_k}]$ should be
redistributed among the weights $w_{jl}$ of the neighbors $l$ of
$j$ according to
\begin{equation}
w_{jl}\rightarrow w_{jl}+\delta'\frac{s_l}{\sum_{k \in
{\cal{V}}(j)}\,s_k}
\end{equation}
where ${\cal{V}}(j)$ indicates the set of neighbors of $j$ and
$k\neq i$.

A detailed computational study on the newly proposed mechanism in
this note will be considered in our future work.

\end{document}